\documentclass[aps,twocolumn,superscriptaddress,groupedaddress,nofootinbib]{revtex4}
\usepackage{graphicx}  
\usepackage{dcolumn}   
\usepackage{bm}        
\usepackage{amssymb,xcolor}   
\usepackage{amsmath}
\usepackage{hyperref} 
\usepackage{ulem}
\usepackage{mwe}
\begin{document}

\title{Enhanced $\psi^{\prime}$ yield and $\psi^{\prime}/(J/\psi)$ yield ratio as a possible signature of QGP formation in high multiplicity $p+p$ collisions}
\author{Partha Bagchi\footnote{parphy@niser.ac.in; parphy85@gmail.com}}

\address{School of Physical Sciences, National Institute of Science Education and Research,
		An OCC of Homi Bhabha National Institute, Jatni-752050, India}

\author{Arpan Das \footnote{arpan.das@pilani.bits-pilani.ac.in}}

\address{Department of Physics,  Birla Institute of Technology and Science Pilani, Pilani Campus, Pilani,  Rajasthan-333031, India}

\author{Ananta P. Mishra\footnote{apmishra@gmail.com}}

\address{School of Physical Sciences, National Institute of Science Education and Research,
		An OCC of Homi Bhabha National Institute, Jatni-752050, India}

\author{Ankit Kumar Panda\footnote{ankitkumar.panda@niser.ac.in}}

\address{School of Physical Sciences, National Institute of Science Education and Research,
		An OCC of Homi Bhabha National Institute, Jatni-752050, India}

\date{\today}

\begin{abstract}
Suppression in the yield of quarkonia (heavy quark-antiquark bound states) has been considered one of the important signatures of the formation of the thermalized deconfined partonic matter, also known as the Quark Gluon Plasma (QGP), in Relativistic Heavy Ion Collision Experiments (RHICE).
Traditionally, the in-medium dissociation of quarkonium states has been presented by implicitly
  assuming an adiabatic approximation, which considers that the heavy quark Hamiltonian
  changes slowly over time owing to change in the medium.
 However, in high multiplicity smaller systems, such as in $p+p$ collisions, the early development of transverse flow resulting from the finite transverse size of the locally thermalized medium may cause the quarkonium states to undergo a non-adiabatic evolution. It has been argued that in the presence of such a non-adiabatic evolution, the suppression of 
 heavy quark-antiquark bound state yields may not reliably indicate QGP formation~\cite{Bagchi:2023vfv}.
We propose that, rather than 
 concentrating on the suppression of $J/\psi$ yields, the enhancement in the yield ratio of $\psi^{\prime}$ to $J/\psi$ (i.e., $\psi^{\prime} / (J/\psi)$), along with an increase in $\psi^{\prime}$ yield, should be considered as a probe of QGP formation for small systems. Our findings, based on realistic modeling of the time evolution of small systems, suggest that the yield ratio $\psi^{\prime} / (J/\psi)$ and the yield of $\psi^{\prime}$ increase as a function of hydrodynamization temperature incorporating the non-adiabatic transitions in high multiplicity $p+p$ collisions. 
\end{abstract}

\maketitle
\section{Introduction}

The suppression in the yield of quarkonia (heavy quark-antiquark bound states), e.g., $J/\psi$ suppression, in heavy ion collision experiments serves as the probe for deconfinement~\cite{PHENIX:2011img,Brambilla:2016wgg,Armesto:2015ioy}. Suppression of quarkonia in a thermalized medium occurs due to the Debye screening of the heavy quark-antiquark potential~\cite{Matsui:1986dk,Kaczmarek:2005ui,Rothkopf:2019ipj}. When the temperature of the deconfined QCD medium becomes higher than the dissociation temperature of these bound states, the quark-antiquark potential becomes fully Debye screened, leading to the dissociation of these states. Traditionally, in much of the existing literature, the dissociation probability of quarkonium states has been derived by assuming an adiabatic evolution of the underlying time-dependent Hamiltonian. Within the adiabatic approximation, it is considered that the heavy quark potential varies slowly over time, implying a gradual change in the Hamiltonian. However, this adiabatic approximation might not always be valid, especially when the Hamiltonian changes rapidly with time. The rapid evolution of the Hamiltonian could be attributed to several factors: the rapid thermalization of the QCD medium~\cite{Bagchi:2014jya}, the fast cooling of the QCD medium~\cite{Dutta:2019ntj}, the quick decay of the time-dependent magnetic field produced in heavy ion collision experiments~\cite{Bagchi:2023jjk}, and the interaction of quarkonia with the chromo-electromagnetic field at $Z(3)$ interfaces in the QCD medium~\cite{Atreya:2014sea}, among others. It is important to note that the difference between the dissociation probability of $J/\psi$ states obtained using adiabatic and non-adiabatic methods can be as large as $36\%$~\cite{Boyd:2019arx}. 

Non-adiabatic evolution of the quarkonia can also arise in small systems, e.g., in $p+p$ collision. Recent experimental data on elliptic flow coefficients for $p+Pb$ system at $\sqrt{s_{NN}} = 5.02$ TeV~\cite{CMS:2014und} and $p+p$ system at $\sqrt{s} = 5~\rm{TeV}$, $7~\rm{TeV}$, and $13~ \rm{TeV}$ at the Large Hadron Collider (LHC)~\cite{CMS:2016fnw,Nagle:2018nvi} indicates the formation of QGP medium in small systems. Unlike the systems produced in the collisions of heavy ions, in $p+p$ collisions, the small transverse size of the system can lead to a rapid decrease in the temperature of the system thereby reducing the lifetime of the deconfined QCD medium.  Moreover, in small systems, a faster change in the system temperature introduces a rapid change in the Hamiltonian entailing a non-adiabatic evolution of bound states of charmonia ($c-\bar{c}$ bound states). Qualitatively, the effect of non-adiabaticity is determined by the interplay of two different time scales. The first one is the time scale corresponding to the change in Hamiltonian ($t_{\rm H}\sim\langle{m}|\dot{H}|n\rangle^{-1} $) and the second one is the time scale associated with the transition to the nearest eigenstate ($t_{\rm tr}\sim |E_m-E_n|^{-1}$)~\cite{Bagchi:2015fry}. If $t_{\rm H}>> t_{\rm tr}$ then the system evolves gradually and the initial eigenstates have sufficient time to adjust to the evolving time-dependent Hamiltonian, which prevents any transition between different eigenstates. We emphasize that the time scale $t_{\rm H}$ crucially depends on the plasma dynamics and is sensitive to the time evolution of temperature.
Rapid temperature evolution of the medium can potentially lead to a situation where the condition for adiabatic evolution, i.e., $t_{\rm H} >> t_{\rm tr}$  may not always be satisfied. Such a scenario may arise in the $p+p$ system where the adiabatic evolution of the bound states may not be used naively.  

Although incorporating the traditional adiabatic evolution quarkonia suppression has been considered a viable signal for QGP formation in $p+p$ collisions \cite{Singh:2021evv,Das:2022lqh}, in a recent study some of us have argued that adiabatic evolution may not be used so naively in small systems as non-adiabatic evolution for small systems can extend the longevity of quark-antiquark bound states, even at higher multiplicities \cite{Bagchi:2023vfv}. We argued that due to non-adiabatic effects, the dissociation probability of $J/\psi$ states has a system size dependence, i.e., for small systems, the dissociation probability of $J/\psi$ can be suppressed, and only for large systems dissociation of $J/\psi$ can be significant. Furthermore, due to the non-adiabatic evolution, the dissociation probability of $J/\psi$ states is not significantly large even if the system temperature is higher than the Debye screening temperature. Since the non-adiabatic evolution raises the question about the viability of $J/\psi$ suppression as a signature of the QGP formation in small systems we lack 
a conclusive baseline for comparing the yield of quarkonia in small systems. As a resolution of this problem and to probe the non-adiabatic evolution of heavy quark-antiquark bound states in this study we propose a possible signature of QGP formation in high multiplicity small systems. We argue that instead of focusing on the suppression of $J/\psi$ yields, the yield ratio of $\psi^{\prime}$ to $J/\psi$ (i.e., $\psi^{\prime}/ (J/\psi)$) along with the increased yield of $\psi^{\prime}$ should be regarded as an indicator of QGP formation. Our approach is similar to the method discussed in Refs.~\cite{Bagchi:2014jya,Dutta:2019ntj,Bagchi:2023jjk}. We utilize the sudden approximation within the context of time-dependent perturbation theory to calculate the dissociation and transition probabilities among various states of $c-\bar{c}$ bound states. In our calculations, the time dependence of the Hamiltonian arises from the time dependence of the system temperature, which we determine using the analytic solution of the (1+1) dimensional Gubser flow and (2+1) dimensional hydrodynamic flow obtained from the MUSIC hydrodynamic code~\cite{Schenke:2010nt,Schenke:2010rr,Paquet:2015lta} with suitable initial conditions. Our findings suggest that the yield ratio $\psi^{\prime}/ (J/\psi)$ and the yield of $\psi^{\prime}$ increases 
 as a function of initial hydrodynamization temperature due to the non-adiabatic evolution of the quarkonia in high multiplicity $p+p$ collisions and can be used as a signature of QGP formation in high multiplicity small systems. It is crucial to note that in certain cases, the adiabatic treatment of quarkonia evolution can lead to an increased yield ratio of $\psi^{\prime}/ (J/\psi)$, as demonstrated in Ref.~\cite{Singh:2021evv}. However, this increase in the yield ratio $\psi^{\prime}/ (J/\psi)$ does not come with an increased yield of $\psi^{\prime}$. In such situations, the yield ratio $\psi^{\prime}/ (J/\psi)$ rises because the survival probability of $J/\psi$ decreases more rapidly than that of $\psi^{\prime}$. Conversely, in the non-adiabatic treatment of quarkonia evolution, we observe a simultaneous increase in both the yield ratio $\psi^{\prime}/ (J/\psi)$ and the yield of $\psi^{\prime}$.

The paper is structured as follows: In Section.\eqref{formalism}, we examine the non-adiabatic approximation within time-dependent perturbation theory to determine the transition probability between various $c-\bar{c}$ states. The non-adiabatic behavior of the Hamiltonian is significantly influenced by the time dependence of the medium temperature. In this section, we also discuss the modeling of temperature evolution in both the pre-hydrodynamic and hydrodynamic stages. Section.\eqref{results} presents the key findings of the study, showcasing the survival probability of different $c-\bar{c}$ states and illustrating that the enhanced ratio of $\psi^{\prime}$ to $J/\psi$ yields along with the enhancement in the yield of $\psi^{\prime}$ can serve as a signature of non-adiabaticity and the formation of a QGP medium in a $p+p$ system. Finally, Section.\eqref{summary} concludes the paper by summarizing the results and offering perspectives for future research.

\section{Formalism}
\label{formalism}
In this section we examine time-dependent perturbation theory, concentrating on the impacts of rapid changes in the Hamiltonian on the time evolution of bound states. Before delving into the detailed framework, we first outline and summarize the key aspects of the time-dependent potential and its implications for quarkonia in a rapidly expanding medium:
\begin{itemize}
    \item The bound state potential for quarkonia is time-dependent due to the sudden change in medium temperature as a result of rapid expansion.
    \item The time dependence in the Hamiltonian arises as the medium temperature evolves over time.
    \item Before hydrodynamization, the medium temperature (effective temperature) is governed by the pre-equilibrium dynamics of the partons.
    \item After hydrodynamization, the temperature information is obtained by solving the Gubser hydrodynamic solution and using the Hydrodynamic model with the MUSIC-Hydro code.
    \item Finally, we solve the time-dependent Schr\"{o}dinger equation perturbatively to determine the survival probability of different bound states.
\end{itemize}

For a detailed discussion, we organize the framework into three main parts. First, we examine time-dependent perturbation theory, focusing on the effects of rapid changes in the Hamiltonian on the time evolution of bound states. Next, we investigate the time evolution of the \textit{effective} temperature during the pre-equilibrium phase. Finally, we assess the time evolution of temperature once the system achieves local thermal equilibrium, using the hydrodynamic approach to model the bulk evolution of the system.

\subsection{Time-dependent perturbation theory: non-adiabatic evolution of quantum states}

Quarkonia are produced during the initial stage of the collision. Since there is no locally thermalized medium produced in the initial stages we can determine the initial state of quarkonia by solving the Schr$\ddot{o}$dinger equation with zero-temperature Hamiltonian $H_0 = \vec{p}^2/2M+\sigma r - \frac{4}{3}\alpha_s/r$~\cite{Wong:1995jf}. Here $M$ represents the reduced mass of the quark-antiquark system. Subsequently, the system thermalizes due to multiple scattering among the partons produced in the initial hard processes. Using the QCD kinetic theory, one can argue that starting from an interacting out-of-equilibrium state, a thermalized medium is attained at a later time denoted as $\tau_{\rm Hydro}$. Once the system achieves thermalization the zero-temperature Hamiltonian evolves into its finite temperature counterpart. Using the KMS model (Karsch-Mehr-Satz model)~\cite{Karsch:1987pv} parametrization of the finite temperature effective potential for quarkonia one can find different $c-\bar{c}$ states by solving the corresponding Schr$\ddot{o}$dinger equation involving the finite temperature Hamiltonian  $H= \vec{p}^2/2M+ \frac{\sigma}{\mu}(1 - \exp(-\mu r)) - \frac{4}{3}\alpha_s \exp(-\mu r)/r$~\cite{Karsch:1987pv}. Here $\alpha_s$ represents the strong coupling constant.  The finite temperature generalization of quarkonia potential has two important parts. The first one is associated with the string tension ($\sigma$), and the second one is associated with the Yukawa term. In the KMS model, the same temperature-dependent Debye screening mass ($\mu = \sqrt{6\pi\alpha_s} T$) appears in both Yukawa and string terms~\cite{Karsch:1987pv,Strickland:2011aa,Dumitru:2009ni}. 
Subsequently, due to the expansion of the system, the medium temperature decreases and eventually, it descends below the quark-hadron transition temperature ($T_c$). Below $T_c$ there is no deconfined medium implying that the in-medium Hamiltonian boils down to the zero temperature Hamiltonian. For small systems at LHC energy scales the thermalization of the system is expected to be very fast within a time scale of the order of $\sim 0.1-1.0$ fm. Moreover in small systems, the subsequent temperature evolution can be fast with respect to heavy-ion collision systems owning to the finite size of the medium and faster transverse expansion. Therefore in small systems the condition for adiabaticity, i.e., $t_{\rm H}>> t_{\rm tr}$ may not be valid and one has to incorporate the non-adiabatic effects~\cite{Bagchi:2023vfv}.

Due to non-adiabatic evolution, the initial quarkonia states undergo transitions to states orthogonal to their initial configurations. Consequently, calculating the probability of the transition of original states
  to these orthogonal states involves a generalization of the time-dependent perturbation theory method, as discussed in  \cite{messiah1961quantum,landau1991quantum,Bagchi:2015fry} for quantum systems. Let us assume at $\tau=0$ the initial state $|i\rangle$ is an eigenstate of the unperturbed Hamiltonian $H_0$, and it evolves to a generic state $|\psi\rangle$ in response to the perturbation $H'(\tau) = H(\tau) - H_0$. Our goal is to find the probability of transition from the initial state to all other states orthogonal to the initial state $|i\rangle$. To accomplish this, we introduce a projection operator $Q = 1 - |i\rangle\langle i|$, designed to project out all states orthogonal to $|i\rangle$. Consequently, any state orthogonal to $|i\rangle$ can be represented as $Q|\psi\rangle$. The transition probability $\mathcal{P}$ is then formulated as follows:
\begin{equation}
\label{eq:zeta1}
 \mathcal{P} = \langle\psi|Q|\psi\rangle.
\end{equation}
Any generic state $|\psi\rangle$ can be written in terms of the eigenstates $|n\rangle$ of the Hamiltonian $H_0$.
\begin{equation}
\label{eq:psi1}
|\psi\rangle = \sum_n c_{ni} |n\rangle.
\end{equation}
The coefficients $c_{ni}$ can be determined using perturbation theory, particularly from first-order perturbation theory considering the states are slowly varying with time (within the first order of perturbations, we do not consider any leading order temporal dependence of states) in comparison to the variation of the perturbation $H'(\tau)$~\cite{Bagchi:2015fry}:
\begin{align}
\label{eq:cnsudn}
 & c_{ni} \,=\, \delta_{ni}-i\langle n|\left(\int_0^{\Delta\tau} 
 H'(\tau)d\tau\right)|i\rangle  \nonumber \\
 &=\, \delta_{ni}- i\langle n|[{H'(\tau)}\tau]_0^{\Delta\tau}|i\rangle +i\langle n|\left(\int_0^{\Delta\tau} \frac{d H'(\tau)}{d\tau}\tau d\tau\right)|i\rangle \nonumber \\
 &=\, \delta_{ni}+i\langle n|\widehat{H}^{\prime}|i\rangle.
\end{align}
Here, we explicitly use the boundary condition for the perturbation, i.e., 
$H^{\prime}(0)=H^{\prime}(\Delta \tau)=0$ and the dimensionless quantity $\widehat{H}^{\prime}$ is defined as,
\begin{equation}
\label{eq:vbar}
 \widehat{H}^{\prime} =\int_0^{\Delta\tau} \frac{d H'(\tau)}{d\tau}\tau d\tau .
\end{equation}
Using Equations.\eqref{eq:zeta1}-\eqref{eq:vbar}, we arrive at the transition probability $\mathcal{P}$~\cite{Bagchi:2015fry}:
\begin{equation}
\label{eq:zeta2}
 \mathcal{P} =  \langle\textbf{} i|\widehat{H}^{\prime}Q\widehat{H}^{\prime}|i\rangle
+\mathcal{O}((\widehat{H}^{\prime})^3).
\end{equation}
It is crucial to emphasize that the accuracy of the coefficient $c_{ni}$ in Equation.\eqref{eq:cnsudn} is limited to the first order of $\widehat{H}^{\prime}$. Consequently, the value of $\mathcal{P}$ is accurate up to the second order in $c_{ni}$, denoted as $(\widehat{H}^{\prime})^2$.
Furthermore, we can express $\langle i|\widehat{H}^{\prime}Q\widehat{H}^{\prime}|i\rangle$ as:
\begin{eqnarray}
 \langle i|\widehat{H}^{\prime}Q\widehat{H}^{\prime}|i\rangle =  \langle i|\widehat{H}^{\prime 2}|i\rangle
 - \langle i|\widehat{H}^{\prime}|i\rangle^2 .
\end{eqnarray}
This allows us to express $\mathcal{P}$ as:
\begin{equation}
\label{eq:zeta-transition}
\mathcal{P} =  \left(\langle i|{\widehat{H}^{\prime 2}}|i\rangle - \langle i|\widehat{H}^{\prime}|i\rangle^2\right).
\end{equation}
As stated above, \(\mathcal{P}\) quantifies the transition probability from the initial state to all states orthogonal to that initial state. Consequently, the survival probability \(\Gamma\) can be expressed as \(\Gamma = 1 - \mathcal{P}\).
Certainly, for the present scenario, we consider the initial state $|i\rangle$ to be $|J/\psi\rangle$, $|\chi_c\rangle$, and $|\psi^{\prime}\rangle$ ($\psi^{\prime}$ is also known as $\psi(2S)$ state). Then the survival probability $\Gamma(J/\psi)$, $\Gamma(\psi^{\prime})$, and $\Gamma(\chi_c)$ represent the survival of those states in the thermalized partonic medium. This is not the full story, there is also the possibility of regeneration of different states due to transition between different states under the influence of non-adiabatic perturbation. Such transitions will modify the survival probability of different initial states. The probability of transition from $|i\rangle$ to $|j\rangle$ is exactly the same as the probability of transition from  $|j\rangle$ to $|i\rangle$ and is equal to $|c_{ij}|^2$, defined in Eq.(\ref{eq:cnsudn}). This will modify the net survival of different states. We quantify the survival of different states as the normalized quantity, 
\begin{align}
\widetilde{\Gamma}_{i} & =\frac{\mathcal{N}_i(1-\mathcal{P}_i)}{\sum_l \mathcal{N}_l}+\frac{\sum_{j}\mathcal{N}_j\times|c_{ij}|^2}{\sum_l \mathcal{N}_l}.
\label{eq:gamma}
\end{align}
Here $\mathcal{N}_i$ denotes the initial population of $|i\rangle$ state. Naturally the initial population of $J/\psi$, $\psi^{\prime}$, and $\chi_c$ states are different~\cite{Singh:2021evv, ALICE:2021qlw}. For convenience, we considered $\widetilde{\Gamma}_{i}$ as the normalized survival probability. If we do not consider the contribution of regeneration because of the transition from other states, then the normalized survival probability can be  expressed as
\begin{equation}\label{eq:gammap}
\widetilde{\Gamma}_{i}^{\prime}=\frac{\mathcal{N}_i(1-\mathcal{P}_i)}{\sum_l \mathcal{N}_l}.
\end{equation}
Using Eqs.\eqref{eq:gamma} and \eqref{eq:gammap} we can calculate the normalized survival probability of different $c-\bar{c}$ bound states once we have the information about the time dependence of the Hamiltonian, i.e., $H^{\prime}(\tau)$, and
$dH^{\prime}/d\tau$. 

\subsection{Modelling of Pre-equilibrium dynamics :}
The time dependence of \(H^{\prime}(\tau)\) and \(\frac{dH^{\prime}}{d\tau}\) stems from the time dependence of the system's temperature. However, modeling the temperature's evolution throughout the entire process is complex. Initially, the temperature is zero at \(\tau = 0\) because no medium has been produced. At a later time, \(\tau = \tau_{\rm Hydro}\), the system reaches local thermal equilibrium, allowing the use of a well-defined hydrodynamic framework to model the temperature's time evolution.

Modeling the temperature evolution for \(0 \leq \tau \leq \tau_{\rm Hydro}\) is particularly challenging because the system is not in local thermal equilibrium during this interval. Nevertheless, for simplicity, we can use the QCD kinetic theory~\cite{Kurkela:2018oqw,Kurkela:2018xxd,Kurkela:2018vqr} to obtain an effective description of the temperature (\(T_{\rm eff}\)) for the pre-equilibrium stage \(0 \leq \tau \leq \tau_{\rm Hydro}\). It is important to note that \(T_{\rm eff}\) is considered an \textit{effective} temperature, as the concept of temperature is strictly defined only in equilibrium. Using the bottom-up thermalization approach discussed in Refs.~\cite{Kurkela:2018oqw,Kurkela:2018xxd,Kurkela:2018vqr}, we adopt the following ansatz for the proper time evolution of \(T_{\rm{eff}}\): 
\begin{equation}
\frac{T_{\rm{eff}}}{T_{\rm{Hydro}}}=\left(\frac{\tau}{\tau_{\rm{Hydro}}}\right)^{\frac{1}{7}\frac{\alpha-1}{(\alpha+3)}}
\label{equ9}
\end{equation}
The parameter \(\alpha\) provides insights into the thermalization dynamics during the pre-equilibrium phase. The underlying physical concept in the above modeling of \(T_{\rm{eff}}\) is that, through microscopic scattering processes, the initially out-of-equilibrium partons become thermalized. The dimensionless \(\mathcal{O}(1)\) parameter \(\alpha\) appearing in the power law determines the rate at which thermalization can be achieved. In principle, \(\alpha\) can take both positive and negative values~\cite{Kurkela:2018oqw,Kurkela:2018xxd,Kurkela:2018vqr}. 

However, we consider only positive values of \(\alpha\) because we are focusing on the initial \textit{heating up}, where \(T_{\rm eff} = 0\) at \(\tau = 0\) and \(T_{\rm eff} \leq T_{\rm Hydro}\) for \(\tau \leq \tau_{\rm Hydro}\). This consideration is important for the survival of the initially produced quarkonia states during the pre-equilibrium stage for \(\tau \leq \tau_{\rm Hydro}\). For positive values of \(\alpha\), the above ansatz immediately implies that at \(\tau = 0\), \(T_{\rm eff} = 0\), and at \(\tau = \tau_{\rm Hydro}\), the effective temperature rises to \(T_{\rm Hydro}\).

It is noteworthy that for certain negative values of \(\alpha\), such as \(\alpha = -1\), one finds that \(T_{\rm eff} \geq T_{\rm Hydro}\) for \(\tau \leq \tau_{\rm Hydro}\). We do not consider such choices in our calculation because, in that case, there may not be any quarkonia states in the medium before the formation of a thermalized partonic medium (for more discussion, see Section II B in Ref.~\cite{Bagchi:2023vfv}). Apart from the power-law rise of \(T_{\rm eff}\), one can also model the initial \textit{heating up} stage where the effective temperature rises linearly from zero to \(T_{\rm eff}\). Both of these parameterizations of \(T_{\rm eff}\) yield qualitatively similar results~\cite{Bagchi:2023vfv}. Such simple ansatz for \(T_{\rm eff}\) approximations can be useful for calculating the average perturbation \(\widehat{H}^{\prime}\) in the pre-thermalization stage.

\begin{figure}
\begin{center}
\includegraphics[width=0.5\textwidth]{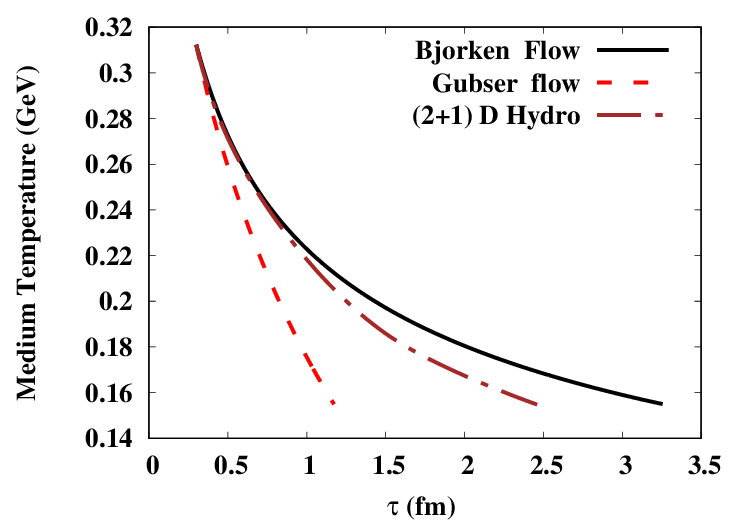}
\caption{The proper time (\(\tau\)) dependence of the medium temperature (\(T\)) results from the hydrodynamic evolution of the thermalized medium. For comparison, we present the temperature evolution for the analytical Bjorken flow (black solid line), the analytical Gubser flow (red dashed line), and the 2+1 D numerical hydrodynamic flow (brown dashed-dotted line). In the Gubser flow solution, we consider \(r_T = 1.5\) fm. For all types of hydrodynamic flow, we assume \(4\pi\eta/s = 1\).}
\label{fig:tau_t_0}
\end{center}
\end{figure}

\subsection{Hydrodynamic modelling :}
As previously mentioned, the formation of a medium in $p+p$ collisions has been the subject of extensive theoretical and experimental investigation. Recent studies indicate the possible formation of a QGP medium in high multiplicity $p+p$ collisions~\cite{Shuryak:2013sra,Song:2017wtw,ALICE:2016fzo,Gutay:2015cba,Jacazio:2024qpb,Bautista:2015kwa,Nagle:2018nvi}. In the current study, we proceed with the assumption that a QGP medium is formed in these systems and undergoes hydrodynamic evolution.

Assuming this scenario, it becomes essential to model the hydrodynamic behavior, especially considering the initial pre-equilibrium dynamics described by the power law formula in Eq.\eqref{equ9}. In our model, quarkonia states form before the establishment of the locally equilibrated deconfined medium, i.e., before the hydrodynamization time scale. As the bulk system evolves, the in-medium quarkonium potential changes accordingly, affecting the overall dynamics of the quarkonia.

We consider two types of hydrodynamic evolution: Gubser flow~\cite{Gubser:2010ui} and a realistic viscous hydrodynamic evolution using the 2+1 D MUSIC code~\cite{Schenke:2010nt,Schenke:2010rr,Paquet:2015lta}. Gubser flow assumes boost-invariant longitudinal expansion and hydrodynamic transverse expansion of a circularly symmetric blob of strongly coupled conformal plasma with a four-velocity $u^{\tau}(\tau,\eta,x_{T})$, independent of the azimuthal angle $\phi$. This approach provides an analytical solution to the equations of the relativistic viscous conformal hydrodynamic system. The Gubser flow gives us the temperature profile of the medium evolution, which can be derived by solving the following set of equations:

\begin{eqnarray}
 \frac{d\hat\epsilon}{d\rho} &=& -\left(\frac{8}{3}\hat\epsilon-\hat{\pi} \right)\tanh(\rho), \label{eq:e_evolution}
\\   \frac{d\hat\pi}{d\rho}  &=& -\frac{\hat\pi}{\hat\tau_{\pi}}+\tanh(\rho)\left(\frac{4}{3}\hat\beta_{\pi}-\hat\lambda\hat\pi- \hat\chi\frac{\hat{\pi}^2}{\hat\beta_{\pi}}\right).\label{eq:pi_evolution}
\end{eqnarray}
Here, key thermodynamic quantities include energy density ($\epsilon$) and shear stress ($\pi$). The dimensionless quantities $\hat{\epsilon}$ and $\hat{\pi}$ are expressed as $\hat{\epsilon}= \hat{T}^4 =\epsilon \tau^4 = 3\hat{P}$ and $\hat{\pi}=\pi\tau^4$, where $\tau$ is the proper time and $\hat{T}$ is related to temperature. The parameter values are chosen as $\epsilon = \frac{3}{\pi^2} T^4$, $\hat{\tau}_\pi = c/\hat{T}$ (related to relaxation time), where $c=5\frac{\eta}{s}$, $\hat{\beta}_\pi= 4\hat{P}/5$, $\hat{\lambda}=46/21$, and the third-order correction parameter $\hat{\chi}=72/245$~\cite{Chattopadhyay:2018apf,Dash:2020zqx}.
The conformal time $\rho$ is given by:
\begin{equation}\label{eq:survival}
\rho = -\sinh^{-1}\left(\frac{1-q^2\tau^2+q^2x_T^2}{2q\tau}\right), 
\end{equation}
where $q$ is an arbitrary energy scale related to the transverse size of the medium ($r_T$) as $q=1/r_T$, and $x_T$ is the position in the transverse plane. The Bjorken flow solution can be retrieved by taking the limit $r_T \rightarrow \infty$ or $q \rightarrow 0$. The temperature evolution for the Gubser flow is shown as a red dashed line in Fig.\eqref{fig:tau_t_0} for $x_T=0$.

Next, we explore a more realistic scenario where the hydrodynamic evolution is modeled by solving the complete set of 2+1 D viscous hydrodynamic equations assuming boost invariance using the well-known MUSIC code~\cite{Schenke:2010nt,Schenke:2010rr,Paquet:2015lta}. However, other hydro packages and formalisms are also available~\cite{Heinz:2005bw,Shen:2014vra,Luzum:2008cw,Chaudhuri:2008sj,Romatschke:2007mq,Bozek:2011ua} that can be used.

In our case, hydrodynamic evolution starts at an early time of 0.3 fm (for both Gubser and 2+1 D viscous hydro cases) with an initial temperature ranging from 1.5 $T_c$ to 4 $T_c$, where $T_c$ is 155 MeV, the hadronization temperature. The realistic hydrodynamic flow is initialized using a smoothly varying Gubser solution of energy density, with subsequent evolution governed by a constant shear viscosity to entropy ratio of $1/4\pi$ and no bulk viscosity included. The evolution stops once the energy density drops below $0.26$ \text{GeV fm}$^{-3}$. The equation of state used here is the lattice EOS hotQCD~\cite{PhysRevD.90.094503}.

This temperature evolution is shown as a brown dashed line in Fig.\eqref{fig:tau_t_0}. Comparing the Gubser and realistic hydrodynamic evolutions used in this study with the Bjorken case (for comparison only) in Fig.\eqref{fig:tau_t_0} indicates that cooling is much faster in the Gubser case. One reason for this could be the transverse size, which is 1.5 fm~\cite{McLerran:2013oju} for Gubser flow, with transverse expansion in the realistic Hydro-code, and infinite transverse size for the extreme Bjorken case. The variation of temperature with system size clearly indicates that for small systems, the time evolution of the system can be rapid compared to large systems, allowing us to explore the scenario of non-adiabatic evolution.

\section{results and discussions}
\label{results}

This section presents significant findings, specifically the normalized survival probability of various charmonium states. It is important to note that transitions among different charmonium states are influenced by the time evolution of temperature, which in turn affects the time evolution of the Hamiltonian. The temperature evolution is sensitive to factors such as $\tau_{\rm Hydro}$, $T_{\rm Hydro}$, the system's transverse size, and transport coefficients like $\eta/s$.
 In our previous work \cite{Bagchi:2023vfv}, we demonstrated that these factors impact the transition probability of $c-\bar{c}$ bound states to unbound states, and we expect similar results here. Among these factors, $T_{\rm Hydro}$ has the most substantial effect on the survival probability of different bound states. Other factors also influence the survival probability, but their impact is relatively minor. We show that as the hydrodynamization temperature ($T_{\rm Hydro}$) increases, the ratio of the normalized survival probabilities of $\psi^{\prime}$ and $J/\psi$ states, i.e., $\widetilde{\Gamma}(\psi^{\prime})/\widetilde{\Gamma}(J/\psi)$, grows.
\footnote{In the present manuscript, we present the variation of the survival probability of different quarkonia bound states with the hydrodynamization temperature, i.e., $T_{\rm Hydro}$. Moreover, in Ref.~\cite{Singh:2021evv}, the authors presented estimates of the survival probability of quarkonia states with the rapidity distribution of charged particles, i.e., $dN_{\rm ch}/d\eta$. Here, $\eta$ should not be confused with the coefficient of shear viscosity. We consider $T_{\rm Hydro}$ as a proxy for $dN_{\rm ch}/d\eta$. Our argument for such a choice is based on Bjorken's estimate of the initial energy density for a heavy ion collision system~\cite{PhysRevD.27.140}. We use the same estimate for $p+p$ collisions as well. According to this approach, a higher $dN_{\rm ch}/d\eta$ implies a higher initial energy density in the collision region, leading to a higher hydrodynamization temperature. Although the exact relationship between initial energy density and system temperature may depend on microscopic details, it is reasonable to infer that greater energy deposition in the central rapidity region will result in a medium with a higher temperature.}

We calculated normalized survival probabilities of $J/\psi$, $\psi^{\prime}$, and $\chi_c$ at different hydrodynamization temperatures, while keeping $4\pi\eta/s=1$, $\tau_{\rm Hydro}=0.3$ fm, and $r_T=1.5$ fm. For the calculation of normalized survival probability, we consider $\mathcal{N}_{J/\psi}:\mathcal{N}_{\psi^{\prime}}:\mathcal{N}_{\chi_c}=100:10:30$~\cite{Singh:2021evv}. We then determine the ratio of normalized survival probabilities of $\psi^{\prime}$ to $J/\psi$.
Moreover, in this study, we assume that quarkonia originate at the center of the medium ($x_T=0$) with no initial transverse momentum $(p_T)$ at $\tau=0$, aiming to maximize the duration within the QGP and thus enhancing the probability of dissociation. We consider $\alpha=2$ to model the rise of the effective temperature ($T_{\rm eff}$) in the pre-equilibrium stage.

\begin{figure}
\begin{center}
\includegraphics[width=0.45\textwidth]{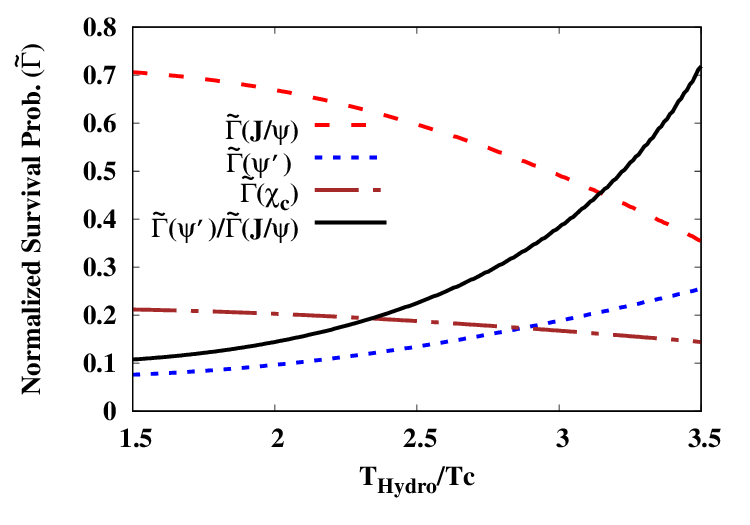}
\caption{Normalized survival probability (including the regeneration effects) of ${J/\psi}$ (red dashed line), ${\psi^{\prime}}$ (blue dotted line), ${\chi_c}$ (brown dash-dotted line) and the ratio of the normalized survival probabilities of ${\psi^{\prime}}~\&~{J/\psi}$ (solid black line) as a function of hydrodynamization temperature in the unit of $T_c$. Here the time evolution of temperature in the hydrodynamic phase has been modelled using the analytical Gubser flow solution. We consider the effect of regeneration due to nonadiabatic transition among different states.}
\label{fig2}
\end{center}
\end{figure}

\begin{figure}
\begin{center}
\includegraphics[width=0.45\textwidth]{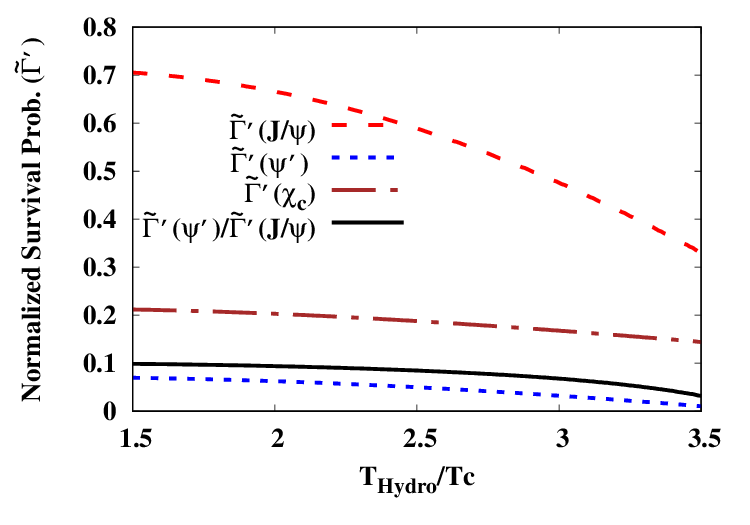}
\caption{Normalized survival probability of ${J/\psi}$ (red dashed line), ${\psi^{\prime}}$ (blue dotted line),  ${\chi_c}$ (brown dash-dotted line) and the ratio of the normalized survival probabilities of ${\psi^{\prime}}~\&~{J/\psi}$ (solid black line) as a function of hydrodynamization temperature in the unit of $T_c$. To model the time evolution of temperature in the hydrodynamic stage we take Gubser flow solution. Here we do not consider the effect of regeneration due to nonadiabatic transition among different states.}
\label{fig3}
\end{center}
\end{figure}

\begin{figure}
\begin{center}
\includegraphics[width=0.45\textwidth]{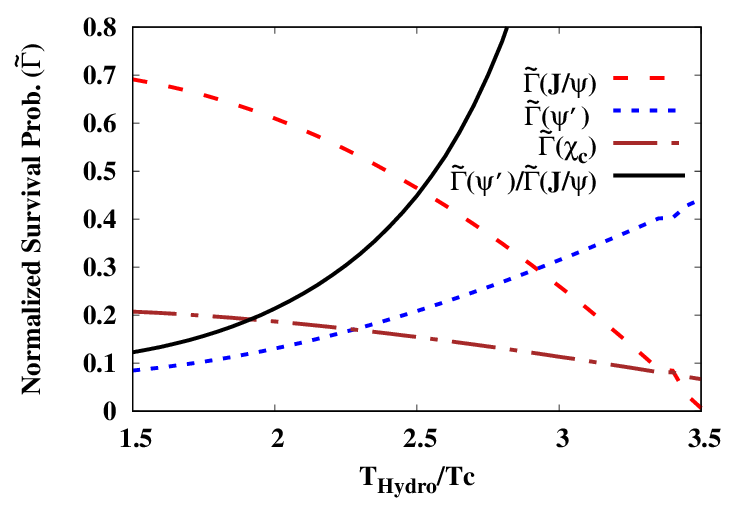}
\caption{Variation of normalized survival probability of ${J/\psi}$ (red dashed line), ${\psi^{\prime}}$ (blue dotted line), ${\chi_c}$ (brown dash-dotted line) and the ratio of normalized survival probabilities of ${\psi^{\prime}}~\&~{J/\psi}$ (solid black line) with $T_{\rm Hydro}/T_c$. Here we model the time evolution of temperature for (2+1) D hydrodynamic evolution using MUSIC hydrodynamic code. We consider the effect of regeneration due to nonadiabatic transition among different states.}
\label{fig4}
\end{center}
\end{figure}

\begin{figure}
\centering
\includegraphics[width=0.45\textwidth]{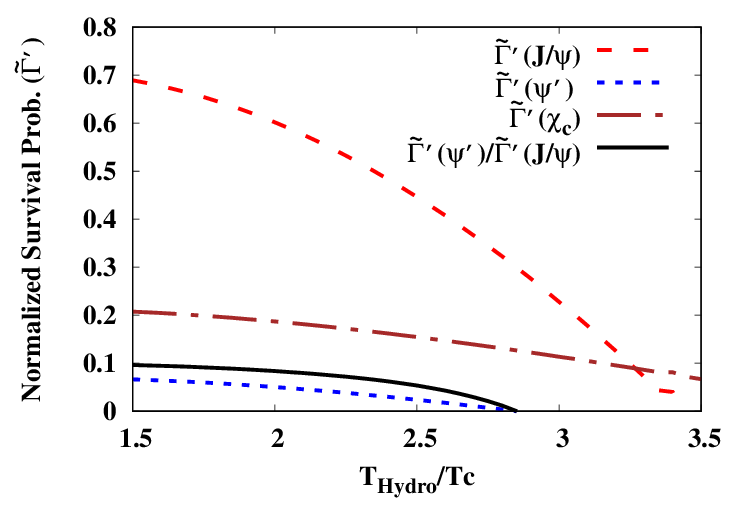}
\caption{Variation of normalized survival probability (without considering the regeneration effects) of ${J/\psi}$ (red dashed line), ${\psi^{\prime}}$ (blue dotted line), ${\chi_c}$ (brown dash-dotted line) and the ratio of the normalized survival probabilities of ${\psi^{\prime}}~\&~{J/\psi}$ (solid black line) with $T_{\rm Hydro}/T_c$. Here (2+1) D hydrodynamic evolution using MUSIC hydrodynamic code has been used to obtain the time evolution of temperature in the hydrodynamic stage.}
\label{fig5}
\end{figure}

In Fig.\eqref{fig2}, we present the normalized survival probabilities of various $c-\bar{c}$ bound states, incorporating the effect of regeneration as defined in Eq.\eqref{eq:gamma}. The red dashed line, blue dotted line, brown dash-dotted line, and solid black line represent the variation of the normalized survival probabilities of $J/\psi$, $\psi^{\prime}$, $\chi_c$, and the ratio of the normalized survival probabilities of $\psi^{\prime}$ to $J/\psi$ with scaled $T_{\rm Hydro}$, respectively. 
In Fig.\eqref{fig3}, we display the same survival probabilities with scaled $T_{\rm Hydro}$, but without considering the regeneration effect as indicated in Eq.\eqref{eq:gammap}. In both Fig.\eqref{fig2} and Fig.\eqref{fig3}, the time evolution of temperature in the hydrodynamic phase is modeled using the analytical Gubser flow solution.

From Fig.\eqref{fig3}, we observe that the normalized survival probabilities of $J/\psi$, $\psi^{\prime}$, and $\chi_c$ decrease with $T_{\rm Hydro}/T_c$ when the effect of regeneration due to non-adiabatic transitions among different $c-\bar{c}$ states is not considered. Additionally, the ratio of the normalized survival probabilities of $\psi^{\prime}$ to $J/\psi$, i.e., $\widetilde{\Gamma}^{\prime}(\psi^{\prime})/\widetilde{\Gamma}^{\prime}(J/\psi)$, also decreases with $T_{\rm Hydro}/T_c$ under the same conditions. 
However, when the regeneration effect due to non-adiabatic transitions is considered, the normalized survival probability of $\psi^{\prime}$ and the ratio $\widetilde{\Gamma}(\psi^{\prime})/\widetilde{\Gamma}(J/\psi)$ are significantly affected. In this case, both $\widetilde{\Gamma}(\psi^{\prime})$ and $\widetilde{\Gamma}(\psi^{\prime})/\widetilde{\Gamma}(J/\psi)$ increase with $T_{\rm Hydro}/T_c$. This can be observed from Figs.\eqref{fig2}.

Conversely, from Figs.\eqref{fig2} and \eqref{fig3}, we observe that the regeneration effect has no significant impact on the normalized survival probability of the $\chi_c$ state. This is because both $J/\psi$ and $\psi^{\prime}$ states are $l=0$ (S-states), while $\chi_c$ is $l=1$ (P-state)~\cite{Deng:2016stx}. As we are only considering spherically symmetric perturbations, the transition probabilities from $J/\psi$ to $\chi_c$ and from $\psi^{\prime}$ to $\chi_c$ are exactly zero, leaving only the transitions between $J/\psi$ and $\psi^{\prime}$ states to be non-zero. 
Although the transition probabilities between $J/\psi$ and $\psi^{\prime}$ are non-zero, their respective initial populations differ by an order of magnitude. Therefore, the normalized survival probability of $\psi^{\prime}$ is significantly more affected by the regeneration effect compared to the normalized survival probability of $J/\psi$. 

It is important to note that in both figures (Fig.\eqref{fig2} and Fig.\eqref{fig3}), the survival probabilities of different bound states remain non-zero even for temperatures higher than the dissociation temperature. This is not expected within the adiabatic treatment of quarkonia evolution.

Finally, in Figs.\eqref{fig4} and \eqref{fig5}, we present the results for the normalized survival probability of different $c-\bar{c}$ states, with the time evolution of the medium temperature modeled using a 2+1 D hydrodynamic evolution. The MUSIC hydrodynamic code was utilized to obtain the proper time evolution of system temperature during the hydrodynamic stage. In Fig.\eqref{fig4}, we show the results incorporating the effect of quarkonia state regeneration. 
The results in Figs.\eqref{fig4} and \eqref{fig5} are qualitatively similar to those in Figs.\eqref{fig2} and \eqref{fig3}, but they differ quantitatively due to the different temperature evolutions in the Gubser flow compared to the realistic 2+1 D hydrodynamic evolution. As observed in Fig.\eqref{fig:tau_t_0}, the system temperature decreases much faster in the Gubser flow than in the 2+1 D hydrodynamic evolution. Therefore, in the 2+1 D hydrodynamic evolution, the system temperature remains above the dissociation temperature for a longer period, contributing to a larger decrease in the survival probability of various $c-\bar{c}$ bound states compared to the Gubser flow. 
This can be observed from Figs.\eqref{fig3} and \eqref{fig5}. This decrease in the individual survival probability is true only as long as we ignore the transitions among different bound states originating from non-adiabatic effects or regeneration effects. For the parameter range we consider, the transition from $J/\psi$ to $\psi^{\prime}$ is higher in the case of 2+1 D hydrodynamics compared to the Gubser flow. This higher regeneration or transition effect results in a higher survival probability of $\psi^{\prime}$ and a higher ratio of the normalized survival probabilities, $\widetilde{\Gamma}^{\prime}(\psi^{\prime})/\widetilde{\Gamma}^{\prime}(J/\psi)$, in the 2+1 D hydrodynamic flow compared to the Gubser flow case.

For completeness, it is worth noting that the authors in Ref.\cite{Singh:2021evv} suggested a potential enhancement in the ratio of the survival probabilities of $\psi^{\prime}$ to $J/\psi$ with increasing charged particle multiplicity through a more detailed modeling of the adiabatic evolution of $c-\bar{c}$ bound states in a QCD medium. However, such an enhancement in the ratio of survival probabilities of $\psi^{\prime}$ to $J/\psi$ is not accompanied by an increase in the survival probability of $\psi^{\prime}$. This is one of the key differences between the results obtained in Ref.\cite{Singh:2021evv} and our present calculation. In our current study, we demonstrate that in the presence of regeneration effects, not only does the ratio of the survival probabilities of $\psi^{\prime}$ to $J/\psi$ increase, but the individual survival probability of $\psi^{\prime}$ also increases.

\section{Summary and outlook}
\label{summary}
In this paper, we estimate the survival probabilities of $J/\psi$, $\psi^{\prime}$, and $\chi_c$ states for small systems using the non-adiabatic evolution of $c-\bar{c}$ bound states. As suggested in Ref.~\cite{Bagchi:2023vfv}, the dissociation probability of quark-antiquark bound states can remain significantly suppressed in small systems, even with high multiplicity, due to non-adiabatic evolution. The qualitative findings reported in Ref.~\cite{Bagchi:2023vfv} indicate that quarkonia suppression might not serve as a definitive signature of deconfinement in high-multiplicity small systems.

To address this issue, we propose that, instead of focusing on quarkonia suppression, the ratio of the survival probability or yield of $\psi^{\prime}$ to $J/\psi$, i.e., $\psi^{\prime}/(J/\psi)$, along with the survival probability of $\psi^{\prime}$, should be considered as a signature of QGP formation. To support this argument, we calculate the normalized survival probabilities of $J/\psi$, $\psi^{\prime}$, and $\chi_c$ under non-adiabatic evolution, which arises from rapid temperature changes in small collision systems. We incorporate non-adiabatic effects using the sudden approximation within the framework of time-dependent perturbation theory to determine the survival probabilities of $J/\psi$, $\psi^{\prime}$, and $\chi_c$.

Our calculations indicate that, due to non-adiabatic evolution, the normalized survival probabilities of $J/\psi$, $\psi^{\prime}$, and $\chi_c$ decrease with the hydrodynamization temperature. However, non-adiabatic evolution can also lead to quantum transitions between different $c-\bar{c}$ bound states, allowing for the regeneration of states from other bound states. By employing both an analytical solution of dissipative hydrodynamics and a more realistic numerical hydrodynamic framework, we demonstrate that the regeneration effect can significantly influence the survival probabilities of different bound states.

We show that as a result of regeneration effects caused by non-adiabatic transition, the survival probability of $\psi^{\prime}$ increases relative to the survival probability of $J/\psi$. This increase in the survival probability of $\psi^{\prime}$ enhances the ratio of the survival probability or yield of $\psi^{\prime}$ to $J/\psi$. The increase in the yield of $\psi^{\prime}$ relative to $J/\psi$ is a consequence of the non-adiabatic evolution in small systems and may serve as a probe for deconfinement. The predicted enhancement in the yield of $\psi^{\prime}$ compared to $J/\psi$ is a testable outcome of our calculations, and its experimental observation could indicate non-adiabatic evolution in small systems.

Several avenues for advancing the current formalism exist. One promising direction is to incorporate the influence of magnetic fields during the medium evolution via MHD formalism~\cite{Gursoy:2018yai,Das:2017qfi,Denicol:2018rbw,Panda:2020zhr,Panda:2021pvq}. Given that magnetic field affect the temperature evolution of the medium, it could potentially alter the survival probabilities significantly. Another important step forward would involve relaxing the longitudinal boost invariance assumption in the hydrodynamic framework and conducting the study using a complete 3+1 dimensional hydrodynamic evolution. However, these considerations lie beyond the scope of the present study and remain topics for future investigation.

\section*{Acknowledgements}
We would like to acknowledge Victor Roy and Nirupam Dutta for fruitful discussions. P.B. acknowledges financial support from DAE project RIN 4001. A. D. acknowledges the New Faculty Seed Grant (NFSG), NFSG/PIL/2024/P3825, provided by the Birla Institute of Technology and Science Pilani, Pilani Campus, India.
A.P. acknowledges the CSIR-HRDG financial support.

\bibliographystyle{apsrev4-1} 
\bibliography{qqbib}{}

\end{document}